# On the possibility to observe relations between quantum measurements and the entropy of phase transitions in $Zn_2(BDC)_2(DABCO)$

Svetlana G. Kozlova[*], Denis P. Pishchur

[1] Nikolaev Institute of Inorganic Chemistry, Siberian Branch, Russian Academy of Sciences, Lavrentyev Av., 3, RU-630090 Novosibirsk, Russian Federation

*Corresponding author. E-mail: sgk@niic.nsc.ru

**Abstract**

The work interprets experimental data for the heat capacity of $Zn_2(BDC)_2(DABCO)$ in the region of second-order phase transitions. The proposed understanding of the processes occurring during phase transitions may be helpful to reveal quantum Zeno effects in metal-organic frameworks (MOFs) with evolving (unstable) structural subsystems and to establish relations between quantum measurements and the entropy of phase transitions.

## 1 Introduction

The evolution of a quantum system can be reduced by repeated contacts with the measuring device. The effect underlying this process is referred to as the quantum Zeno effect (QZE) [1-10]. The evolution of a quantum system can be also accelerated by contacts with a measuring device. The corresponding effect is referred to as the quantum anti-Zeno effect (QAZE) [11-13]. However, quantum measurements cannot be completely isolated from the surrounding environment in real-world conditions. The evolution of the tested system can be suppressed or accelerated depending on the character of the system's Hamiltonian that describes its interaction with the environment; therefore, QZE/QAZE effects may be caused by the action of periodically changing interactions in the environment rather than only by a series of measurements [12-17]. Of particular interest is the problem related to the establishment of thermal equilibrium between a quantum system and its environment. In spite of significant progress that has been achieved in quantum thermodynamics in recent years, some fundamental problems still remain open. Most models imply weak interactions between the test system and the environment, but weak interactions mean that such systems would not be fairly effective in practice [18]. In this paper, we consider an interaction between quantum measurements and the entropy of phase transitions in the $Zn_2(BDC)_2(DABCO)$ metal-organic framework (MOF). This interaction may promote the development of thermodynamic control of QZE and QAZE in solid-phase systems.

## 2 Conditions for observing QZE and QAZE effects in $Zn_2(BDC)_2(DABCO)$

### 2.1 The system to detect and study QZE and QAZE effects

The QZE and QAZE effects are detected and studied using the $Zn_2(BDC)_2(DABCO)$ system containing organic ligands $BDC^{2-}$ (anion of terephthalic acid $(C_8H_4O_4)^{2-}$) and DABCO (1,4-diazabicyclooctane molecule $C_6H_{12}N_2$) (Fig.1); the framework of this system can absorb various molecules, including atomic gases [19, 20]. Since it was previously shown that the DABCO molecule has two energy degenerate quantum states corresponding to right-twisted (R) and left-twisted (S) structural modifications [21-28], the ensemble of DABCO molecules is viewed here as an unstable subsystem. The specific heat of $Zn_2(BDC)_2(DABCO)$ corresponds to the 1D one-dimensional vibrational continuum and exhibits three second-order phase transitions (PTs) and [26]. PT1 is associated with the ordering/disordering of $BDC^{2-}$ anions at a phase transition point $T_c \sim 130K$ [20]. PT2 ($T_c \sim 60K$) is associated with the termination of tunnelling between two energy-degenerate quantum states (R and S) of DABCO molecules; PT3 ($T_c \sim 14K$) is related to the distribution of DABCO molecules over three different energy states [21-25, 28]. Of primary interest are low-temperature phase transitions PT2 and PT3 associated with the evolution of DABCO energy states and, consequently, with a possible occurrence of QZE/QAZE effects.

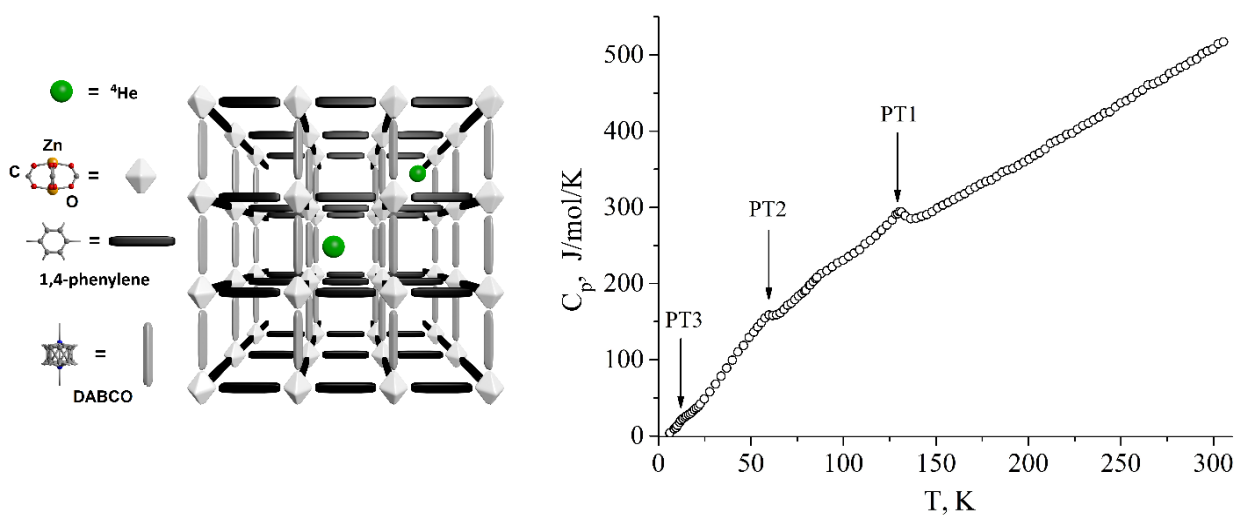

**Fig. 1** Schematic structure of $Zn_2(BDC)_2(DABCO)$ whose framework is characterized by a substantial volume of free space (more than 60% of the framework volume). Temperature dependence of heat capacity ($C_p$).

## 2.2 "Measuring device"

The suggested "measuring device" is helium-4 atoms (Fig. 1) appearing in structural positions next to DABCO molecules in $Zn_2(BDC)_2(DABCO)$. Such localization is substantiated by theoretical studies aimed at finding preferred positions of xenon atoms in a similar MOF $Ni_2(2,6\text{-}ndc)_2(DABCO)$. It was shown that the most energetically profitable position for xenon atoms is close to DABCO [29]. It is known that helium-4 atoms in liquid and solid phases are characterized by the vibration frequency of ~ $10^{13}$ Hz and exhibit a large amplitude of zero vibrations (~1 Å), which may be even larger in $Zn_2(BDC)_2(DABCO)$ pores. Therefore, DABCO molecules will be periodically affected by the neighboring helium-4 atoms to possibly cause QZE or QAZE effects. Obviously, the condition required for the realization of QZE/QAZE effects is $\nu \geq \nu_a$, where $\nu$ is the vibration frequency of helium and $\nu_a$ is the energy barrier (in frequency units) between two DABCO states. This condition is satisfied, since $\nu \sim 10^{13}$ Hz and $\nu_a \approx 1.3 \cdot 10^{12}$ Hz (for PT2 at $T_c \sim 60K$) and $\nu_a \approx 0.3 \cdot 10^{12}$ Hz (for PT3 at $T_c \sim 15K$). The inverse time $\nu_{eq}$ required for the establishment of thermal equilibrium between evolving DABCO molecules and the $Zn_2(BDC)_2(DABCO)$ matrix depends on the interaction of DABCO molecules with the environment; its value is unknown but is approximately estimated to be ~ $10^{14}$ Hz (the interaction energy between organic linkers and metal atoms in the MOF was estimated as ~ 100 kJ/mol [29]). Thus, our conditions for the observation of QZE/QAZE are determined by relations $\nu_{eq} \geq \nu \geq \nu_a$ (i.e., QZE/QAZE will be observed if the interaction between the evolving system and its environment is not weak).

## 2.3 Approach to detect QZE or QAZE

The QZE/QAZE can be discovered by the changes in the entropy and the temperature of phase transitions in $Zn_2(BDC)_2(DABCO)$. This statement is substantiated by the properties of order-disorder phase transitions. If only some part of particles (in our case, evolving DABCO molecules) interact with vibrating helium atoms, the entropy of phase transitions will be changed. In the case of QZE, the entropy will decrease, since some particles do not participate in the phase transition. In the limiting case, when the "measuring device" registers all particles, the phase transition will be either inhibited completely or occur at higher temperatures as the sample is heated. In the case of QAZE, the picture should be opposite, and in the limiting case when all particles are "measured", the phase transition will occur at lower temperatures as the sample is heated. The changes in entropy can be found by analyzing the heat capacity of phase transitions (PT2 and PT3) at various concentrations ($n$) of absorbed helium-4 atoms in the pores of $Zn_2(BDC)_2(DABCO)$. The presence of quantum coherence in the studied system is a necessary condition for the Zeno effect. In our case, this condition means that the system preserves its critical state in the region of phase transitions (i.e., phase transitions associated with the absorption of helium atoms into the MOF pores must be second-order phase transitions).

**2.4 Experimental results**

Temperature dependence of heat capacity of $Zn_2(BDC)_2(DABCO)$ in the region of 4.2-300 K have been repeatedly measured. The first results were obtained in [26], the values of molar entropy ($\Delta S_m$) associated with phase transitions were analyzed in detail in [30]. The Table summarizes the results of determining $\Delta S_m$ values for order-disorder phase transitions in $Zn_2(BDC)_2(DABCO)$ depending on the gas pressure (P) in the calorimeter measuring cell. According to our estimations, the concentration $n$ under these pressures varied from 0.50 to 1.50 atoms per 10 unit cells of $Zn_2(BDC)_2(DABCO)$. As can be seen, $\Delta S_m$ values depend on P during PT2 and PT3 (related to the evolution of DABCO molecules) and do not depend on P during PT1 (associated with the ordering/disordering of $BDC^{2-}$ anions).

**Table** Entropy ($\Delta S_m/R$) of phase transitions in the region of critical temperatures under various pressures (P, Pa) of the heat-exchange gas ($^4$He) normalized against molecular weight of organic fragments in $Zn_2(BDC)_2(DABCO)$. R is the universal gas constant.

| P | $\Delta S_m/R$ (PT1) | $\Delta S_m/R$ (PT2) | $\Delta S_m/R$ (PT3) |
|---|---|---|---|
| $0.51 \cdot 10^5$ | 0.53±0.06 | 0.7±0.1 | 2.2±0.2 |
| $1.52 \cdot 10^5$ | 0.53±0.06 | 1.2±0.1 | 1.4±0.2 |

**3 Discussion of obtained results and conclusion**

According to the results obtained in [30], the increase in the number of helium-4 atoms in $Zn_2(BDC)_2(DABCO)$ pores (and, therefore, in the number of contacts between DABCO molecules and helium-4 atoms) led to the changes in $\Delta S_m$ values of phase transitions PT2 and PT3. In line with the proposed approach (section 2.3), it can be assumed that the QZE and the QAZE will occur during PT3 and PT2, respectively. Thus, the sorption of helium-4 atoms in the pores of $Zn_2(BDC)_2(DABCO)$ accelerates the termination of transitions between R- and S-quantum states of DABCO during PT2; on the contrary, the redistribution of DABCO over different states during PT3 is slowed down. Note that the occurring phase transitions are still second-order phase transition, and coherent interactions between the particles of the crystal are preserved [30]. Interestingly, ordering and disordering of $BDC^{2-}$ anions is not affected by absorbed helium-4 atoms during PT1. We explain this by the fact that the structure of DABCO, in contrast to that of $BDC^{2-}$ anions, is flexible (see theoretical calculations [31,32]) so that helium-4 atoms may cause some deformation of DABCO and affect the evolution processes while the structure of $C_8H_4O_4^{2-}$ anions remains unchanged.

The presented analysis shows that $Zn_2(BDC)_2(DABCO)$ and related MOFs containing evolving structural fragments can be interesting experimental systems to study thermodynamic processes occurring in the conditions of the QZE and the QAZE.

All authors contributed equally to the paper.